\newcommand{\be}{\begin{equation}}
\newcommand{\ee}{\end{equation}}
\newcommand{\bea}{\begin{eqnarray}}
\newcommand{\eea}{\end{eqnarray}}
\newcommand{\ba}{\begin{array}}
\newcommand{\ea}{\end{array}}

\newcommand{\ben}{\begin{eqnarray}\displaystyle}
\newcommand{\een}{\end{eqnarray}}

\newcommand{\al}{\alpha}
\newcommand{\ga}{\gamma}
\newcommand{\de}{\delta}
\newcommand{\ep}{\epsilon}

\newcommand{\La}{\Lambda}

\newcommand{\diag}{{\rm diag}}
\newcommand{\D}{{\rm d}}
\newcommand{\pa}{\partial}
\newcommand{\rar}{\rightarrow}
\newcommand{\non}{\nonumber}

\newcommand{\cN}{\mathcal{N}}
\newcommand{\cF}{\mathcal{F}}
\newcommand{\cO}{\mathcal{O}}

\newcommand{\PP}{\mathrm{I}\kern -2.5pt \mathrm{P}}
\newcommand{\R}{\mathrm{I}\kern -2.5pt \mathrm{R}}
\newcommand{\Z}{\mathsf{Z}\kern -5pt \mathsf{Z}}
\newcommand{\sR}{\mathrm{I}\kern -1.5pt \mathrm{R}}
\newcommand{\sC}{\mathsf{C}\kern -4.1pt \mathsf{I}}
\newcommand{\bint}{\int\kern -12.5pt -}

\newcommand{\sectiono}[1]{\section{#1}\setcounter{equation}{0}}

\newcommand{\half}{{\textstyle {1\over 2}}}
\newcommand{\ts}{\textstyle}

\newcommand{\1}{1\kern -3pt \mathrm{l}}

\newcommand{\SU}{\mathrm{SU}}
\newcommand{\SO}{\mathrm{SO}}
\newcommand{\Sp}{\mathrm{Sp}}
\newcommand{\U}{\mathrm{U}}

\documentclass[12pt]{article}

\usepackage{epsfig}
\usepackage{amssymb}

\newdimen\tableauside\tableauside=1.0ex
\newdimen\tableaurule\tableaurule=0.4pt
\newdimen\tableaustep
\def\phantomhrule#1{\hbox{\vbox to0pt{\hrule height\tableaurule
width#1\vss}}}
\def\phantomvrule#1{\vbox{\hbox to0pt{\vrule width\tableaurule
height#1\hss}}}
\def\sqr{\vbox{%
  \phantomhrule\tableaustep

\hbox{\phantomvrule\tableaustep\kern\tableaustep\phantomvrule\tableaustep}%
  \hbox{\vbox{\phantomhrule\tableauside}\kern-\tableaurule}}}
\def\squares#1{\hbox{\count0=#1\noindent\loop\sqr
  \advance\count0 by-1 \ifnum\count0>0\repeat}}
\def\tableau#1{\vcenter{\offinterlineskip
  \tableaustep=\tableauside\advance\tableaustep by-\tableaurule
  \kern\normallineskip\hbox
    {\kern\normallineskip\vbox
      {\gettableau#1 0 }%
     \kern\normallineskip\kern\tableaurule}%
  \kern\normallineskip\kern\tableaurule}}
\def\gettableau#1 {\ifnum#1=0\let\next=\null\else
  \squares{#1}\let\next=\gettableau\fi\next}

\newcommand{\Ysymm}{\tableau{2}}
\newcommand{\Yasymm}{\tableau{1 1}}

\begin{document}

\begin{flushright}
{\tt hep-th/0404125}\\
CERN-PH-TH/2004-068 \\
\end{flushright}
\vspace{1mm}
\begin{center}
{\bf\Large\sf
A note on instanton counting for {\large $\cN=2$} gauge theories \\
with classical gauge groups }
\end{center}
\vskip 5mm
\begin{center}
Marcos Mari\~{n}o\footnote{Also at Departamento de 
Matem\'atica, IST, Lisboa, Portugal} and Niclas Wyllard
\end{center}

\begin{center}
Department of Physics, CERN, Theory Division, 1211 Geneva 23, Switzerland \\[3mm]
{\tt marcos,wyllard@mail.cern.ch}
\end{center}

\vskip 2mm

\begin{abstract}
We study the prepotential of $\cN=2$ gauge theories
using the instanton counting techniques introduced by Nekrasov.
For the $\SO$ theories without matter we find a closed expression
for the full prepotential and its string theory gravitational corrections.
For the more subtle case of $\Sp$ theories without matter we discuss
general features and compute the prepotential
up to instanton number three. We also briefly discuss SU theories
with matter in the symmetric and antisymmetric representations. We
check all our results against the predictions of the corresponding Seiberg-Witten
geometries.
\end{abstract}

\setcounter{equation}{0}
\section{Introduction}

The celebrated Seiberg-Witten solution \cite{Seiberg:1994a}
of $\cN=2$ gauge theories has been studied in
great detail, but until recently no tractable method was known for obtaining
the instanton expansion of the prepotential directly from many-instanton calculus.
The final stumbling blocks were overcome in \cite{Nekrasov:2002}, building
on previous work by several authors (see \cite{Dorey:2002}
for a review of many-instanton calculus predating \cite{Nekrasov:2002}).
Instanton corrections to the prepotential are determined by an integral
over the moduli space of instantons. The crucial idea in~\cite{Nekrasov:2002} was
to use localization techniques in a clever way to show
that the integral over the moduli space of instantons can be computed
from contributions of isolated fixed points, or equivalently, can be
recast as a contour integral. This and related contour
integrals had actually made an
appearance much earlier \cite{Losev:1997}, but
the precise connection between these contour integrals
and the prepotential was established only in \cite{Nekrasov:2002}.
For the gauge theories discussed in \cite{Nekrasov:2002}
the contour integral was explicitly evaluated and the result written as
a sum over partitions. One surprising fact
is that this result also encodes information about all the higher genus
gravitational corrections which appear when the gauge theory is embedded in type II
string theory. The results for these gravitational corrections were
tested using topological strings
in~\cite{Klemm:2002,Iqbal:2003a}.
Further work inspired by these instanton counting techniques can be
found in~\cite{Flume:2002}-\cite{Flume:2004}.

The calculations in \cite{Nekrasov:2002} were done for the gauge group $\SU(N)$.
In this note we also consider the other classical gauge groups, $\SO(N)$ and $\Sp(2N)$.
For the $\cN=2$ $\SO(N)$ gauge theory without matter we find the complete
solution to the instanton counting problem, and write an explicit formula
for the prepotential
and its gravitational corrections as a sum over partitions, as was done
in \cite{Nekrasov:2002} for $\SU(N)$. For the $\Sp(2N)$ theory, we write down
the appropriate contour integral and evaluate it explicitly up to instanton
number three. The structure of the poles for the Sp integral
turns out to be much more complicated than for the cases of SU and SO gauge groups.
We find the locations of all poles of the integrand, but have not been able
to determine in closed form which ones are picked out by the contour prescription
for arbitrary instanton numbers.

For the $\SU(N)$ theory with one hypermultiplet in the symmetric representation
we also find the complete
solution to the instanton counting problem in terms of a sum over partitions.
On the other hand, for the $\SU(N)$ theory with
one hypermultiplet in the antisymmetric representation
we encounter problems similar to the ones that occur in the $\Sp(2N)$ theory.

In retrospect, the fact that $\Sp(2N)$ as well as $\SU(N)$ with antisymmetric matter
are more
complicated is not too surprising. Indeed, these theories are known to be
more subtle already at the level of the Seiberg-Witten
solution \cite{D'Hoker:1997b,Naculich:1998,Ennes:1999}. More recently, these
theories were studied using the Dijkgraaf-Vafa matrix-model
approach~\cite{Dijkgraaf:2002} and a modification of
the original proposal was found to be required for
these models~\cite{Kraus:2003,Naculich:2003b}.

The next section contains the analysis of the instanton counting for
SO and Sp gauge groups. Explicit results for the prepotential up to
instanton number 3 are listed in appendices A and B for SO and Sp,
respectively. Section 3 presents the analysis for SU theories with
matter hypermultiplets in the symmetric or antisymmetric representation, and
explicit results up to instanton number 2 are presented in appendix C.
In section 4 we list some open problems.

\setcounter{equation}{0}
\section{Instanton counting for {\large $\cN=2$ $\SO$/$\Sp$} theories}

In this section we discuss the $\cN =2$ $\SO$/$\Sp$ gauge theories.
After briefly discussing the features common to both models we
will discuss each case in turn.

\subsection{ADHM data}

A particularly expedient way of obtaining the ADHM instanton
equations \cite{Atiyah:1978} for the $\cN=2$ $\SU(N)$ gauge theory is
from a system of $k$ D(-1) branes (where $k$ is the instanton number) and  $N$
D3-branes \cite{Witten:1996}.
The (bosonic) ADHM data can be assembled into the four
complex quantities $(B_1,B_2,I,J)$ where $B_1$ and $B_2$ are in
the adjoint of $\U(k)$, and $I$ and $J$ belong to the bifundamental
representations $(\mathbf{k},\bar{\mathbf{N}})$ and $(\bar{\mathbf{k}},\mathbf{N})$
of $\U(k){\times}\SU(N)$, respectively. The $\U(k)$ gauge symmetry acts as
\be
\label{gaug}
(B_a, I,J) \rightarrow (g B_a g^{-1}, g I, J g^{-1}) \,,
\ee
where $g \in \U(k)$. The ADHM equations are (see e.g.~\cite{Flume:2002} for
more details)
\ben
\mu_{\sR}&=& [B_1, B_1^{\dagger}]+[B_2, B_2^{\dagger}]+I I^{\dagger} -J^{\dagger}J=0\,,
\nonumber\\
\mu_{\sC}& =& [B_1, B_2] + I J =0\,.
\label{adhme}
\een
The moduli space of instantons with instanton number $k$ is the
space of solutions to the above equations, modulo gauge transformations.
As discussed in \cite{Nekrasov:2002},
it is actually convenient to consider a deformation of the first equation in
(\ref{adhme}) and take $\mu_{\sR}\not=0$. The space of solutions to (\ref{adhme})
for $\mu_{\sR}\not=0$ (modulo gauge transformations)
gives a resolution of singularities of the ADHM moduli space
of instantons and can be regarded as the space of solutions to the second
equation modulo {\it complex} gauge
transformations, provided a stability condition is imposed (as usual
in geometric invariant theory).

Notice that the linearization of the gauge symmetry action (\ref{gaug}) gives
a map
\be
\label{mapC}
C: {\bf g} \rightarrow E\oplus E \oplus (V\otimes W^*) \oplus (V^*\otimes W) \,,
\ee
where ${\bf g}$ is the complexified Lie algebra of the instanton symmetry group $U(k)$,
$E$ is the representation space
associated to the matrices $B_a$ (in this case, since $B_a$ belong to
the adjoint representation, $E$ is the complexified
Lie algebra), $V$ is the defining vector space of the instanton symmetry group, 
and $W$ is the defining vector space of the gauge group. One also has
the linearization of the second equation in (\ref{adhme}), which gives a map
\be
\label{map}
s:   E\oplus E \oplus (V\otimes W^*) \oplus (V^*\otimes W)  \rightarrow {\bf g} \,.
\ee
The maps (\ref{mapC}), (\ref{map}) fit together in
the instanton deformation complex
\be
\label{idc}
{\bf g} \stackrel{C}{\rightarrow} E\oplus E \oplus (V\otimes W^*) \oplus (V^*\otimes W) \stackrel{s}{\rightarrow} {\bf g} \,.
\ee
The tangent space to the instanton moduli space is given
locally by ${\rm Ker}\, s /{\rm Im}\,C$, see e.g.~\cite{Dorey:2002,
Nakajima:1999} for more detailed discussions.

The ADHM data for $\SO$/$\Sp$ gauge groups can be obtained by a projection of
the $\SU$ data. The ADHM equations for $\SO$/$\Sp$ were first
obtained in \cite{Corrigan:1978}.

In the D-brane language the projection is implemented by the addition of an
orientifold (O3) plane to the D(-1)/D3 system. When placed on top of the
stack of D3-branes the orientifold plane does not break
any further supersymmetry. Depending
on the charge of the orientifold one obtains either $\SO(N)$ or $\Sp(2N)$ as the gauge
group on the D3's. Due to the properties of the orientifold
projection~\cite{Gimon:1996} one gets the ``opposite'' gauge group on
the D(-1)'s i.e. $\Sp(2k)$ and $\SO(k)$, respectively.

Implementing the projection on the ADHM data shows that for the $\SO(N)$
($\Sp(2N)$) theory, the $B_a$'s are {\it not} in the
adjoint representation, but rather in the other two-index
representation. For $\Sp(2k) $ the adjoint is isomorphic to the symmetric
representation whereas the $B_a$'s are in the antisymmetric.
For $\SO(k)$ the adjoint is isomorphic to the antisymmetric representation whereas
the $B_a$'s are in the symmetric.  The orientifold
projection on $I$ and $J$ relate them to their complex conjugates and thus halves
the number of components.

The instanton deformation complex for $\SO$/$\Sp$ has the
same form as above (\ref{idc}), with the only difference that now $E$
is the appropriate representation space. Notice that this
description gives the correct number of parameters:
the moduli space of instantons on $\R^4$ for a gauge group $G$ has complex dimension
$2k g^{\vee}$, where $g^{\vee}$ is the dual Coxeter number of the group.
For $\SO(N)$ at instanton number $k$, $I$ and $J$
provide $2k N$ complex parameters, while $B_{1,2}$ are in the antisymmetric and give
$2(k^2-k)$ complex parameters. The number of complex parameters modulo gauge transformations and the ADHM constraints is therefore $2k(N-2)$ from which it follows
that the moduli space of instantons has complex dimension $2k(N-2)$.
For $\Sp(2N)$, a similar counting
gives $2k(N+1)$, both in agreement with the general formula for the dimension.

\subsection{{\large Instanton counting for $\SO(N)$}} \label{sSO}

The instanton corrections to the Seiberg-Witten prepotential can be computed
as integrals over the moduli space of instantons; see \cite{Dorey:2002} for
a review. These integrals are difficult to evaluate, but as
shown in \cite{Nekrasov:2002} one can use powerful localization
techniques to simplify their computation\footnote{See
\cite{Hollowood:2002a} for earlier ideas on using
localization to evaluate the moduli space integrals. }.
The localization is done with
respect to the group $\U(1)^N{\times} \U(1)^2$, where
$\U(1)^N$ is the Cartan subgroup of the gauge symmetry, and $\U(1)^2$ is a global
symmetry corresponding to an $\SO(2){\times} \SO(2)$ rotation in spacetime,
i.e.~$\R^4$.
This symmetry acts as follows on the ADHM fields: $(t_1, t_2) (B_1, B_2, I, J) =
(t_1B_1, t_2 B_2, I, t_1t_2 J)$, where $(t_1, t_2) \in \U(1){\times} \U(1)$.
The ADHM equations (\ref{adhme}) are
unchanged under this action. It turns out that the fixed loci of this action
are points,
and this allows one to compute the integrals as sums over contributions coming
from the
fixed points.

Alternatively, one can start with the twisted version of the ${\cal N}=2$ theory
written in terms of the ADHM fields and consider an
equivariant extension of the BRST symmetry with respect to the above group action.
Since the action is BRST exact, the path integral of the twisted theory
can be calculated
in the semiclassical approximation. By integrating out $(B_1,B_2,I,J)$ (and their
fermionic partners) one can reduce the path integral to a much simpler
contour integral over the eigenvalues of a field $\phi$ which is part of the ${\cal N}=2$ topological
multiplet (see \cite{Moore:1998,Bruzzo:2002,Flume:2002} for details on this). From
this contour integral, obtained from the path integral of the twisted theory,
one can in the end extract the prepotential of the original ${\cal N}=2$ theory.
It turns out that the poles that contribute to the contour integral are located precisely at
the fixed points of the $\U(1)^N {\times} \U(1)^2$ action on the instanton moduli space, and the
residue of the integral at each pole is precisely the contribution of the corresponding
fixed point in the localization formula.

The analysis of \cite{Nekrasov:2002} extends to the other classical 
gauge groups, and
one can in fact write down the general form of the resulting
contour integral\footnote{See \cite{Braverman:2004} for the generalization to
general groups of other aspects
of \cite{Nekrasov:2002,Losev:2003,Nekrasov:2003}.}.  As in the $\U(N)$ case, one
has to consider an (equivariant) extension with respect to $H {\times} \U(1)^2$,
where $H$ is the Cartan subgroup
of the gauge group. The field $\phi$ lives
in the adjoint representation of the instanton symmetry group $G_k$ at
instanton number $k$,
which is $\U(k)$, $\Sp(2k)$ and $\SO(k)$ for the gauge groups $\U(N)$,
$\SO(N)$ and $\Sp(2N)$, respectively.
After diagonalization $\phi$ can
be written as a vector ${\bf \phi}=\sum_I \phi_I e_I$ in the root
lattice of $G_k$, where $e_I$, $I=1, \cdots, r_k$,
is an orthonormal basis and $r_k$ is the rank of $G_k$.
Let ${\bf \alpha} \in \Delta$ be the roots of the
instanton symmetry group, and let ${\bf \mu} \in \Lambda_B$ be the weights of the
representation of the instanton
symmetry group where the matrices $B_a$
live. Finally, let $\hat a$ be a vector in the Cartan subalgebra ${\bf h}$ of the
gauge group, and
let $P(\phi_I)= \prod_i (\phi_I-\hat a_i)$. Notice that the $\hat a_i$'s play the
role of equivariant parameters with respect to the symmetry action $H$. The integral then reads:
\be
\label{generalintegral}
Z_k \propto {1\over |{\cal W}_k|}\frac{1}{(2\pi i)^{r_k}}\oint \prod_I {d\phi_I \over P(\phi_I) P(\phi_I + \epsilon)} {\prod_{{\bf \alpha} \in \Delta}
({\bf \phi}\cdot {\bf \alpha})({\bf \phi}\cdot {\bf \alpha} +
\epsilon) \over \prod_{{\bf \mu} \in \Lambda_B} ({\bf \phi}\cdot {\bf \mu} + \epsilon_1)
({\bf \phi}\cdot {\bf \mu} + \epsilon_2)}\,.
\ee
In this formula $\ep=\ep_1{+}\ep_2$, where $\epsilon_{1,2}$ are the equivariant
parameters in the Cartan subalgebra of the $\SO(2){\times} \SO(2)$ rotation
(in other words, $t_i=e^{\epsilon_i}$), and $|{\cal W}_k|$ is the order of the
Weyl group of the instanton symmetry group. In (\ref{generalintegral})
we have omitted an overall factor which depends on $\epsilon_{1,2}$.
This integral has a nice geometric interpretation in terms of the
instanton deformation complex associated to the ADHM equations (\ref{idc}).
This complex has an equivariant extension with respect to the
$H {\times} \U(1)^2$ action. Let $Q$ be the defining representation space for $\U(1)^2$, and let $W$ be the
defining representation for $H$. Then, $(B_1, B_2) \in E \otimes Q$, $I \in
V\otimes W^*$ and $J \in V^*\otimes W \otimes\wedge^2 Q$, and we obtain the equivariant complex:
\be
\label{eqcomplex}
{\bf g} \rightarrow (E\otimes Q) \oplus (V\otimes W^*) \oplus (V^* \otimes W \otimes \wedge^2 Q)
\rightarrow {\bf g}\otimes \wedge^2 Q\,.
\ee
The integrand in (\ref{generalintegral}) then computes the virtual
Euler characteristic of the
complex (\ref{eqcomplex}) \cite{Moore:1998,Nekrasov:2002}. The
denominator corresponds to the middle term, while the numerator corresponds
to the first and third terms.

Let us now consider the ${\cal N}=2$ $\SO(N)$ super Yang-Mills theory.
From the analysis of the
ADHM data we know that the instanton symmetry group is $\Sp(2k)$ and that the fields $B_a$ live
in the antisymmetric representation of $\Sp(2k)$. The contour integral (\ref{generalintegral}) reads
in this case\footnote{When integrating out $J$ one seems to get
$\sqrt{P(\phi_I+\ep)P(-\phi_I+\ep)}$ rather than $P(\phi_I+\ep)$.
However as we exclusively set $\ep_1=-\ep_2$ after evaluating the integral,
this difference does not affect our results. When dealing with the $\ep\neq 0$
expressions this should be kept in mind. (Similar comments apply to the other
contour integrals appearing in this paper.)}
\bea
\label{sonintegral}
Z_k &= &{1\over 2^k k!} {\epsilon^k\over (\epsilon_1 \epsilon_2)^k}
\frac{1}{(2\pi i)^k} \oint \prod_{I=1}^k \D\phi_I
{ (2\phi_I)^2 ((2\phi_I)^2-\epsilon^2)
\over P(\phi_I) P(\phi_I + \epsilon)}  \\
&\times & \!\!\! \prod_{I<J}
\frac{(\phi_I-\phi_J)^2 ((\phi_I-\phi_J)^2-\ep^2) (\phi_I +\phi_J)^2
((\phi_I+\phi_J)^2-\ep^2)}
{((\phi_I-\phi_J)^2-\ep_1^2)((\phi_I-\phi_J)^2-\ep_2^2)
((\phi_I+\phi_J)^2-\ep_1^2)((\phi_I+\phi_J)^2-\ep_2^2)}\,.
\nonumber
\eea
For $\SO(2N)$,
$P(\phi)=\prod_{i=1}^{N} (\phi^2 - a_i^2)$ whereas for $\SO(2N+1)$,
$P(\phi)=\phi \prod_{i=1}^{N} (\phi^2 - a_i^2)$.
Notice that the integrand in (\ref{sonintegral}) is
invariant under permutations of the $\phi_I$'s (which is a group of order
$k!$) and also under the group $\Z_2^m$ generated by
$a_i \rightarrow -a_i$, $i=1, \cdots, m$. This is of course nothing but the
Weyl group of $\Sp(2k)$.

The integral (\ref{sonintegral}) can be evaluated by computing residues at the appropriate poles,
as in \cite{Moore:1998,Nekrasov:2002,Bruzzo:2002}.
In the case of the $\SO(2N)$ integral
(\ref{sonintegral}), the poles turn out to be
essentially the same as the ones for
$\SU(N)$. However, as this is not completely obvious we will give some details.
For $\SU(N)$ the location of the fixed points (and consequently the location 
of the poles) can be determined
by solving the equations
\be
\ba{lll}
[\phi,B_1]=\ep_1 B_1  \,,&  [\phi,B_2] = \ep_2 B_2  \,, &  [B_1,B_2] + IJ = 0\,, \\
\phi I - I \hat{a} = 0 \,,& J\phi - \hat{a} J - \ep J =0\,,&
\ea
\label{fixedeqs}
\ee
modulo complex gauge transformations.
It is important to note that because of the specific choice of integration
contour not all solutions to
these equations are actually relevant to the evaluation of the integral. 
This is natural
if we recall that the solutions to $[B_1,B_2] + IJ = 0$ only give solutions
to the original ADHM equations (deformed to $\mu_{\sR}\not=0$)
provided a certain stability condition is satisfied. It turns out that the poles that
contribute to the
contour integral are in one-to-one correspondence with the solutions of (\ref{fixedeqs})
that satisfy the stability condition. This condition can be phrased as follows. As discussed
above $I \in V\otimes W^*$ where $V$ and $W$ are the representation spaces
for the fundamental representation of $\U(k)$ and $\SU(N)$ respectively. The
solutions which contribute to the integral are the ones which satisfy the
condition that $B_1^{n} B_{2}^{m} I$ for $m,n=1,2,\ldots$ span the vector space $V$.

In \cite{Nekrasov:2002} it was shown that the solution to the equations
(\ref{fixedeqs}) subject to the stability condition are classified by
Young tableaux in the following way.  Each solution at instanton number $k$ is given
by a set of $N$ Young tableaux ${\bf Y} = \{Y_1,\ldots,Y_{N}\}$ subject to the
constraint that $\sum k_{\ell}=k$ where $k_{\ell}$ is the total number of boxes
in the $\ell$th tableau. The boxes in $Y_{\ell}$ are labelled by pairs of
integers $(i_{\ell}, j_{\ell})$, where $i_{\ell}$ labels the rows
and $j_{\ell}$ the columns. In this language the fixed points are given by
\be
\phi_{I_{\ell}} =a_{\ell} -(j_{\ell}-1) \epsilon_1 -(i_{\ell}-1)\epsilon_2\,.
\label{fixedson}
\ee

We will now discuss how one can obtain the solution to the $\SO(2N)$ problem by
imposing a projection on the solution for $\SU(2N)$.
That one gets solutions by this projection is clear, but it is by no means
 a priori obvious that one gets all (relevant) solutions this way. The reason why
there might be a problem is that one does not start with the most general
solution to the $\SU$ problem, but instead only considers the solutions which satisfy
the stability condition.
In fact, as we will see later, for $\Sp(2N)$ this naive method does not
give all solutions. Nevertheless, the simplified method works for $\SO(N)$ and
it is instructive to give some details.
The methods we use (and the equations we solve)
are similar to the ones used  in $\cite{Naculich:2001}$ to derive all solutions
to the classical vacuum equations for $N=1^*$ theories with $\SO/\Sp$ gauge
groups\footnote{In that paper the solutions were characterized in terms of $\SU(2)$
representations. One could also use this language here, but
it is more convenient to work with the Young tableaux.}. After a change of basis to
bring $\phi$ into diagonal form the $\Sp(2k)$ conditions on the matrices
$\phi$, $B_1$ and $B_2$ can be written
\be
\ba{lll}
\phi^T g = -g \phi   \,, \quad B_1^T g = g B_1  \,, \quad  B_2^T g = g B_2  \,, \quad
g^T=-g \, ,\quad gg^*=-\1_{2k} \,.&
\ea
\label{soproj}
\ee
If, given a (stable) solution to the $\SU(2N)$ problem, 
a $2k{\times} 2k$ matrix $g$ can be found such that these equations
are satisfied then the solution will descend to a solution of the
$\SO(2N)$ problem.

In matrix notation the solution to (\ref{fixedeqs}) for $\SU(2k)$ schematically
takes the block diagonal form
\be
\phi = \diag( y_1, ..., y_{2k} ) \,, \quad
B_1 = \diag( x_1, ..., x_{2k} ) \,, \quad
B_2 = \diag( z_1, ..., z_{2k} )\,.
\ee
Here $y_\ell$ is a diagonal matrix with elements $a_\ell - (i_\ell-1) \ep_1 -
(j_\ell-1) \ep_2$  with some ordering for the
Young tableau elements ($i_\ell$ and $j_\ell$ label rows and columns as already
discussed), for instance left-to-right and down-up.
Below we will use that there is a $\Z_2 {\times} \Z_2$ symmetry acting within
{\it each} block,
\bea
&& \ep_1 \rar -\ep_1 \,, \quad X_\ell \rar (X_\ell)^T \,, \non \\
&& \ep_2 \rar -\ep_2 \,, \quad Z_\ell \rar (Z_\ell)^T\,.
\label{Z2Z2}
\eea
(This follows from the form of the equations $[\phi,B_{a}] = \ep_{a} B_{a}$.)

From the $\SO$ condition we get $a_{N+i} = -a_i$, $i=1,\ldots,N$.
The $\Sp(2k)$ condition on $\phi$ can be written $(\phi_I + \phi_J) g_{IJ} = 0$
where $I,J=1,\ldots,2k$. In order for $g_{IJ}\neq 0$ to be possible $\phi_I + \phi_J$
must vanish. For generic $a_{\ell}$, $\ep_1$, and $\ep_2$ this can only happen if $\phi_I$
is in block $\ell$ and $\phi_J$ is in block $\ell+N$ (since then $a_\ell + a_{\ell+N} = 0$).
At first sight it seems that only for $i=j=1$ in both the $\ell$th and the $(\ell+N)$th
block is $\phi_I + \phi_J$ zero. This would contradict the $g^*g = -\1_{2k}$
condition and imply that there are no solutions. However, note that one can
utilize the symmetry (\ref{Z2Z2}) to change the sign of $\ep_{1,2}$ in block $\ell+N$,
$\ell=1,...,N$. After this change, the allowed $g$ takes the form
\be
g = \left(\ba{cc} 0 & D \\ -D & 0 \ea \right)\,,
\ee
where $D = \diag(D_1,...,D_N)$ and each $D_i$ is a diagonal matrix. It is also required
that there is the same number of elements in blocks $i$ and $i+N$ otherwise some
of the $D_i$'s would be rectangular matrices with at least one row (or column)
full of zeroes. This would imply that $\det(g)=0$ which contradicts $g^* g = -\1_{2k}$.

The above implies that the tableaux have to come in pairs; in the $\phi$
matrix one has to have $y_{i+N} = -y_i$. We are not quite finished since
we also have to check that the $\Sp(2k)$
conditions on $B_{1,2}$ can be satisfied. The $\Sp(2k)$ condition on $B_{1}$
becomes (in matrix notation):
$x_i D_i = D_i (x_{i+N})^T $ (and similarly for $z_i$). We can choose a
basis in which all the non-zero elements in $B_1$ are normalized to 1, which
implies $x_i = (x_{i+N})^T$ from which it can be shown that $D_i$ is proportional
to the unit matrix.
The condition $g^* g =-\1_{2k}$ then implies that the proportionality constant is
a phase, which can be removed by a unitary transformation. One also finds
$z_i = (z_{i+N})^T $.

To summarize: we have shown that the (stable) solutions 
to (\ref{fixedeqs}), (\ref{soproj})
(and hence the locations of the contributing poles)
are classified by Young diagrams. Notice that only $\phi_I$ for $I=1,\ldots,k$
enter in the integral. However, apart from the solution in (\ref{fixedson})
there are also contributing solutions that are obtained by permuting
the $k$ $\phi_I$'s, and we also have the possibility of
choosing $\pm \phi_I$. This gives an overall factor $2^k k!$ 
which cancels the normalization $1/|{\cal W}_k|$ in (\ref{sonintegral}).

For $\SO(2N+1)$ the only difference compared to $\SO(2N)$ is that there is now an
extra $\hat{a}_i$, $\hat{a}_0\equiv0$. The matrix $\phi$ still belongs to the adjoint
of $\Sp(2k)$ and the conditions (\ref{soproj}) are unchanged. This means that
there is an extra block that is independent of the $a_i$'s. It not hard to show
that a projection of the solution for $\SU(2N+1)$ would force this block to be
filled with zeroes, but such solutions will always have moduli (giving rise
to a vanishing integral) therefore this block has to be empty/absent.
The solution is thus the same as for $\SO(2N)$.
Note that even though the contributing fixed points are the same the residues
are of course different.

Using a generalization of the method in \cite{Nakajima:1999,Nekrasov:2002} one may
explicitly evaluate the integral in terms of an infinite product. We will only
write the expressions for $\ep_2=-\ep_1=\hbar$, which is enough to extract the Seiberg-Witten
prepotential and the gravitational corrections. The computation can be easily done if
we take into account that the residue at a given pole of the
integral (\ref{generalintegral})
can be translated into a localization computation: first, we
compute the weights of the group action
$H {\times} \U(1)^2$ on the different spaces appearing
in (\ref{eqcomplex}). This is easily done
by using the fact that (\ref{fixedson}) gives the weights associated to the defining
representation of the instanton symmetry group. The residue is then given by the product of the weights
associated to the first and third terms of (\ref{eqcomplex}), divided by the product of the
weights associated to the middle term of (\ref{eqcomplex}).

For a specific partition ${\bf k} = (k_1,\ldots,k_N)$ one finds for $\SO(2N)$
\bea
\label{sonprod}
Z_{\bf k} &=& 16^k\prod_{(i,j)\in Y_{\ell}}
[a_{\ell} + \hbar(i - j - \half)]
[a_{\ell} + \hbar(i - j)]^2 [a_{\ell} + \hbar(i - j + \half)]   \\
& \times &\!\!\! \prod_{Y_{\ell}, Y_{r}} \prod_{m,n=1}^{\infty} \!
\frac{ [a_{\ell} - a_{r} + \hbar(k_{\ell,m} {-} k_{r,n} {+}n {-} m)]
[a_{\ell} + a_{r} + \hbar(k_{\ell,m} {-} \tilde{k}_{r,n} {+}n {-} m)] }
{[a_{\ell} - a_{r} + \hbar(n {-} m)][a_{\ell} + a_{r} + \hbar(n {-} m)]} \,,\non
\eea
where it is assumed that the points where $(l,m)=(r,n)$
are excluded from the product. In (\ref{sonprod}) $k_{\ell,m}$ denotes the number
of boxes in the $m$th row of the $\ell$th tableau and $\tilde{k}_{\ell,n}$
denotes the number of boxes in the $n$th column of the $\ell$th tableau. These
definitions extend to all positive integers $n,m$:
$k_{\ell,m}$ and $\tilde{k}_{\ell,n}$ are defined to be identically zero when
$n,m$ lie outside the tableau.

For $\SO(2N+1)$ one finds the same product as for $\SO(2N)$ with the only difference that the product over single Young tableaux is instead given by
\be
\prod_{(i,j)\in Y_{\ell}}
[a_{\ell} + \hbar(i - j - \half)][a_{\ell} + \hbar(i - j + \half)]\,.
\ee
The partition function for $\SO(N)$ at instanton number $k$, $Z_k$, is obtained
by summing over all possible partitions ${\bf k}$ with $k$ boxes in total, $Z_{k} = \sum_{\bf k} Z_{\bf k}$,
where the sum is over $\bf k$ such that $\sum_{l,m} k_{l,m}=k$.
Finally, the complete partition function is $Z=  \sum_k L^{k}Z_k$ with 
$L=\La^{2N-4}$.
One has the important relation \cite{Nekrasov:2002}
\be
Z = \exp\bigg[ \frac{1}{\ep_1\ep_2} \cF(a,\ep_1,\ep_2,\La) \bigg] \,,
\ee
where (for $\ep_2=-\ep_1=\hbar$)
\be
\cF(a,\hbar,\La) = \sum_{g=0}^{\infty} \hbar^{2g} \cF_g(a,\La) \,,
\ee
with
\be
\cF_g(a,\La) = \sum_{k=1}^{\infty}  L^k F_{g,k}(a) \,.
\ee
Here $\cF_g(a,\La)$ determines the genus $g$ gravitational coupling, and $F_{g,k}$
is the contribution to this coupling at instanton level $k$. In particular, $\cF_0$
is the Seiberg-Witten prepotential (excluding the perturbative part 
which will be discussed below).

We have checked that the above results lead to expressions for the first
three instanton corrections to the prepotential which are in agreement with
the results obtained using the Seiberg-Witten approach.
Details about this check are included in appendix A. One can also check
the results for the gravitational correction ${\cal F}_1$ against the 
general expression
in terms of Seiberg-Witten data derived in \cite{Klemm:2002,Dijkgraaf:2002b} (which is
easily seen to extend to the $SO(N)$ case). This expression involves the determinant 
of the period matrix, whose instanton expansion in powers of $\Lambda^{2N-2}$ 
is better calculated with the techniques 
of \cite{emm} after rewriting it in terms of hyperelliptic theta functions. Using 
this procedure we have checked  the instanton counting result
that $F_{1,1}=0$ for $SO(N)$.

It is also of interest to connect the above discussion more directly to the 
corresponding Seiberg-Witten geometry. In \cite{Nekrasov:2003} it was shown 
how to obtain the Seiberg-Witten data (i.e.~the curve and the differential) 
from the instanton counting results for $\SU(N)$. 
(Another way to check the equivalence with the Seiberg-Witten approach 
was presented in \cite{Nakajima:2003a}; see also \cite{Nakajima:2003b}.)

Here we will briefly discuss how the analysis in \cite{Nekrasov:2003} 
is modified for the case of $\SO(N)$. Following \cite{Nekrasov:2003} we introduce 
\bea
\!\!\!\!  f_{k_\ell}(x,\hbar) &\!\!\!=&\!\! 
|x| + \sum_{i=1}^{\infty} \bigg[ |x {-} \hbar(k_{\ell,i}{-}i{+}1)|
{-}|x{-}\hbar(k_{\ell,i}{-}i)|{+}|x{+}\hbar(i{-}1){+}|x{+}\hbar\, i|\bigg]  \\
&=& \!\! |x| + \sum_{j=1}^{\infty} \bigg[ |x {+} \hbar(\tilde{k}_{\ell,j}{-}j{+}1)|
{-}|x{+}\hbar(\tilde{k}_{\ell,j}{-}j)|{+}|x{-}\hbar(j{-}1)|{+}|x{-}\hbar j|\bigg]\,, \non
\eea
as well as the function $\ga_{\hbar}(x,\La)$ satisfying
\be
\ga_{\hbar}(x+\hbar,\La) + \ga_{\hbar}(x-\hbar,\La) - 2 \ga_{\hbar}(x,\La) =\ln\bigg(\frac{x}{\La}\bigg) \,.
\ee
We furthermore define $f_{{\bf k},a}(x,\hbar) = \sum_\ell f_{k_\ell}(x-a_\ell)$. 
Using these definitions together with $\frac{\D^2}{\D x^2} |x| = 2\de(x)$ 
one can show that 
\bea \label{fullZ}
\mathcal{Z}_{\bf k} &\!\!=&\!\! \exp\bigg\{ -{\ts \frac{1}{4}} \bint f_{{\bf k},a}''(x)f''_{{\bf k},a}(y)\ga_{\hbar}(x-y,\La)
-{\ts \frac{1}{4}} \int  f_{{\bf k},a}''(x)f''_{{\bf k},a}(y)\ga_{\hbar}(x+y,\La) \non \\ &&
\qquad \; + \, {\ts \frac{1}{2}} \int f_{{\bf k},a}''(x) [ 2 \ga_{\hbar}(x,\La) + \ga_{\hbar}(x+\hbar/2,\La) + \ga_{\hbar}(x-\hbar/2,\La) ] \bigg\}
\eea
is equal to $Z_{\mathrm{pert}} L^k Z_{\bf k}$ where $Z_{\bf k}$ was given in (\ref{sonprod}) and 
\bea \label{pertZ}
Z_{\mathrm{pert}} &\!\!=&\!\! \exp\bigg\{ -\sum_{\ell\neq r}\ga_{\hbar}(a_\ell-a_r,\La)
-\sum_{\ell, r} \ga_{\hbar}(a_\ell+a_r,\La) \non \\ && \qquad\, +\, \sum_{\ell} 
[ 2 \ga_{\hbar}(a_\ell,\La) + \ga_{\hbar}(a_{\ell}+\hbar/2,\La) + \ga_{\hbar}(a_\ell-\hbar/2,\La) ] 
\bigg\} \,.
\eea
These expressions are valid for $\SO(2N)$; for $\SO(2N+1)$ the $2\ga_{\hbar}(x,\La)$ 
terms in (\ref{fullZ}) and (\ref{pertZ}) are absent.

Defining $\cF^{\mathrm{pert}}$ via $Z_{\mathrm{pert}} = \exp[\cF^{\mathrm{pert}}/\hbar^2] $ and using \cite{Nekrasov:2003,Nakajima:2003b}
\be \label{ga0}
\hbar^2 \ga_{\hbar}(x,\La) = \frac{x^2}{2} \ln\bigg(\frac{x}{\La}\bigg) 
- \frac{3x^2}{4} + \cO(\hbar^2) \equiv \ga_{0}(x,\La) + \cO(\hbar^2)\,,
\ee
it is easy to check that $\cF^{\mathrm{pert}}_{0}$ agrees with the usual perturbative 
result for the prepotential
\be
 \cF^{\mathrm{pert}}_{0} = - \sum_{\al\in\Delta_{+}} (a \cdot \al)^2 
\ln\bigg(\frac{a\cdot\al}{\La}\bigg) + \mathrm{quadratic}\,,
\ee
where the sum is over the positive roots of $\SO(N)$. The higher order terms 
in $\cF^{\mathrm{pert}}$ give conjectural expressions for the perturbative part 
of the gravitational corrections.

More generally one can analyze the full partition function, $\mathcal{Z}$, obtained by summing over terms of the form (\ref{fullZ}) for all possible partitions at all instanton numbers, in the limit $\hbar\rar0$.  It was argued in \cite{Nekrasov:2003} that in this limit the sum goes over to an integral and $\mathcal{Z}$ can be obtained from the saddle-point of the action ($\mathcal{Z} = \exp[\mathcal{E}/\hbar^2]$)
\be \label{E}
\mathcal{E} =  -{\ts \frac{1}{4}} \bint f''(x)f''(y)\ga_{0}(x{-}y,\La)
-{\ts \frac{1}{4}} \! \int \!\! f''(x)f''(y)\ga_{0}(x{+}y,\La)  +  
2 \!\! \int \!\! f''(x)  \ga_{0}(x,\La) \,,
\ee
where $f''(x)$ is  a continuous (real) function localized on intervals around the $a_i$'s and $\ga_{0}(x,\La)$ was defined in (\ref{ga0}). Comparing this expression to the one in \cite{Nekrasov:2003} we note the expected connection with the $\SU(2N)$ theory with 4 massless hypermultiplets in the fundamental representation. We also note that at the saddle point (\ref{E}) is equal to the total prepotential (including the perturbative piece), $\mathcal{E} = \mathcal{F}_{0, \mathrm{tot}}$. As usual in saddle-point problems of this kind it is convenient to introduce the resolvent 
\be
R(z) = \half \int \D x \, \frac{f''_{a}(x)}{z-x} \,, 
\qquad R(x+i\varepsilon)-R(x-i\varepsilon) = \pi i f_{a}''(x) \,,
\ee
which satisfies 
\be
a_i = \frac{1}{2\pi i } \oint_{A_i} z R(z) \, \D z \,, \qquad
\frac{1}{2\pi i }  \oint_{A_i} R(z) \, \D z = 1 \,,
\ee
where $A_i$ is the contour surrounding the $i$th cut.
It should also be straightforward to derive 
\be
\frac{\pa \cF_{0,\mathrm{tot}} }{\pa a_i} = \frac{1}{2\pi i } 
\oint_{B_i} z R(z) \, \D z \,, 
\qquad \frac{1}{2\pi i } \oint_{B_i} R(z) \, \D z = 0\,,
\ee
and discuss in more detail the form of the curve following \cite{Nekrasov:2003} (see 
also~\cite{Hollowood:2003}) but we will not do so here.

\subsection{Instanton counting for {\large $\Sp(2N)$}}
\label{sSp}

For the $\Sp(2N)$ case, $\phi$ belongs to the adjoint of $\SO(k)$ (where $k$ is
the instanton number) and we have to
distinguish between odd and even instanton numbers. For $k = 2n{+}1$ we have
(using a convenient normalization)
\bea
Z_{2n+1} &=& \frac{(-1)^n}{2^{n+1} n!} \frac{\ep^n}{(\ep_1 \ep_2)^{n+1}} \frac{1}{\sqrt{P(0)P(\ep)} } \frac{1}{(2\pi i)^n} \oint \prod_{I=1}^{n} \D\phi_I
\frac{1}{P(\phi_I) P(\phi_I+\ep)} \non \\ & \times &
\frac{\phi_I^2 (\phi_I^2-\ep^2)}{(\phi_I^2-\ep_1^2)((2\phi_I)^2 -\ep_1^2)
(\phi_I^2-\ep_2^2)((2\phi_I)^2 -\ep_2^2) }  \\
&\times & \!\!\! \prod_{I<J}
\frac{(\phi_I-\phi_J)^2 ((\phi_I-\phi_J)^2-\ep^2) (\phi_I +\phi_J)^2
((\phi_I+\phi_J)^2-\ep^2)}
{((\phi_I-\phi_J)^2-\ep_1^2)((\phi_I-\phi_J)^2-\ep_2^2)
((\phi_I+\phi_J)^2-\ep_1^2)((\phi_I+\phi_J)^2-\ep_2^2)}
\nonumber
\eea
whereas for $k = 2n$ we instead get
\bea
Z_{2n} &=&\frac{(-1)^n}{2^n n!}\frac{\ep^n}{(\ep_1\ep_2)^n} \frac{1}{(2\pi i)^n}
\oint \prod_{I=1}^{n} \D\phi_I
\frac{1}{P(\phi_I) P(\phi_I+\ep)} \frac{1}{((2\phi_I)^2-\ep_1^2)((2\phi_I)^2-\ep_2^2))}
\non \\
&\times & \!\!\! \prod_{I<J}
\frac{(\phi_I-\phi_J)^2 ((\phi_I-\phi_J)^2-\ep^2) (\phi_I +\phi_J)^2
((\phi_I+\phi_J)^2-\ep^2)}
{((\phi_I{-}\phi_J)^2-\ep_1^2)((\phi_I{-}\phi_J)^2-\ep_2^2)
((\phi_I{+}\phi_J)^2-\ep_1^2)((\phi_I{+}\phi_J)^2-\ep_2^2)}.
\eea
One can follow the same approach as for $\SO(N)$ and project
Nekrasov's solution. However, this procedure does {\it not} give the right result:
some solutions are missing. The problem is that some solutions to the fixed point 
equations for $\SU(2N)$, which do not satisfy the stability
condition turn out to contribute (after projection) to the evaluation of the
above integrals. This can be phrased as saying that the stability condition
is different in the $\SU$ and $\Sp$ cases.
 The new stable solutions not obtained via a projection of Nekrasov's solution
 are all of the form $\phi_I = 0 +\cO(\ep_1,\ep_2)$ i.e.~localized near
$\phi_I=0$. These extra solutions are the manifestation in the present framework
of the extra cut around $x=0$ in the Seiberg-Witten curve (\ref{Spcurve})
or equivalently of the ``$\Sp(0)$''
factors~\cite{Kraus:2003,Naculich:2003b}
in the Dijkgraaf-Vafa matrix model approach.

To get an understanding of the extra solutions, we should consider all solutions to
the fixed point equations (\ref{fixedeqs}) together with the $\SO(k)$ projections
\be
\ba{lll}
\phi^T f = -f \phi   \,, \quad B_1^T f = f B_1  \,, \quad  B_2^T f = f B_2  \,, \quad
f^T=f \, ,\quad ff^*= \1_{k} \,.&
\ea
\ee
If, given a solution to (\ref{fixedeqs}), we can find 
a $k{\times}k$ matrix $f$ such that these conditions
are satisfied then the solution to (\ref{fixedeqs}) descend to a solution
of the $\Sp(2N)$ $k$-instanton problem.
We have solved these equations from first principles using the techniques
in~\cite{Naculich:2001}. Some of the solutions have
moduli, i.e.~undetermined parameters. These solutions do not contribute since
they give a vanishing result for the path integral (see e.g.~\cite{Moore:1998}
for a discussion of this point in a similar context). Since the new solutions 
not obtained by projection do not depend on the $a_i$'s we focus on the
$a_i$-independent solutions (such solutions have $I=J=0$). We have found
that these solutions can be represented pictorially in terms of stacks of rows of
boxes. Each box corresponds to an eigenvalue of $\phi$. These boxes are placed in the
$xy$-plane in a symmetric way. The rows are parallel to the $x$-axis and
can be stacked on top of each other. The rules are:
\begin{itemize}
\item The diagram has to be symmetric under $x\rar -x$ together with $y\rar -y$.
\item A row on top of another (above the $x$-axis) can not extend to the right
of the row below it.
\item Rows with one box are special. Such rows can only be placed to the far
left of the row below it.
\item A row of length $n_1$ placed on top of a row of length $n_2$ is not allowed
if $(n_2-n_1)/2$ is equal to the distance between the right end of the lower row
and the right end of the upper row.
\end{itemize}
In general one has several disconnected diagrams. But no two diagrams
are allowed to be the same (this would give moduli). These rules give
$a_i$-independent solutions to the fixed-point equations which do not have moduli.
This is best illustrated by an example. The allowed connected diagrams with
four boxes are
\begin{figure}[h]
\centering
\includegraphics{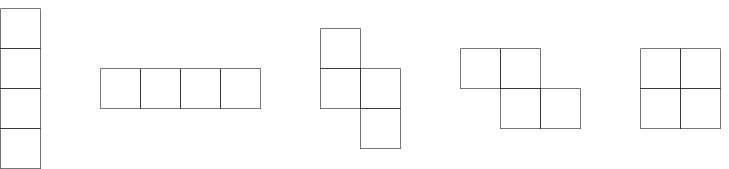}
\center{\small {\sf Figure 1:} 
Connected diagrams relevant to the $k=4$ calculation.}
\end{figure}

\noindent These diagrams correspond to the 
solutions: $(\phi_1,\phi_2) = (\ep_2/2,3\ep_2/2)$;
$(\ep_1/2,3\ep_1/2)$; $(-\ep_1/2,-\ep_1/2+\ep_2)$; $(-\ep_1+\ep_2/2,\ep_2/2)$, $(-\ep_1/2 + \ep_2/2,\ep_1/2+\ep_2/2)$ (modulo the action of the Weyl group). Here $(\phi_1,\phi_2)$ are the two eigenvalues which enter in the integral (the other two components encoded in the above diagrams are minus these by the $\SO$ projection). The Weyl group acts on these solutions and one would expect that the number of times each type of solution appears in the evaluation of the integral to be  equal to an integer multiple of the number of elements in a Weyl orbit, but explicit calculations indicate that this is not the case.

From the above analysis we know all possible locations of the fixed points.
The remaining problem is to determine which of these possibilities are
actually realized in the evaluation of the integral and how many times each
solution appears. In general, not all
solutions are relevant since the choice of integral contour/pole prescription
excludes some possibilities. (It turns out that all the diagrams in the above figure
contribute in the evaluation of the $k=4$ integral.) It is natural to expect
that diagrams
which do not look like two Young diagrams glued together (some with the first
row (column) of half the height (width)) are to be excluded.
There may be some connection between these
diagrams with boxes of half the normal height (width)
and the spinor representations of
$\SO$. In addition to the fact that some diagrams are excluded one also needs to
determine how many times each of the diagrams that do contribute occur.
Unfortunately we have been unable to solve this combinatorial problem,
i.e.~we have not been able to determine what the correct `stability condition'
is for the extra solutions appearing in the $\Sp(2N)$ case.

However, even without an explicit product formula one may still check that the above
integrals lead to expressions which agree with previous results.
In appendix B we check that the above integrals give rise to expressions
for the first three instanton corrections to the prepotential which agree with
the ones obtained using the Seiberg-Witten approach.

\setcounter{equation}{0}
\section{Instanton counting for {\large $\SU(N)$} with $\protect\Ysymm$/$\protect\Yasymm$ matter}

Other examples which can be treated using the methods of \cite{Nekrasov:2002}
include $\cN=2$ $\SU(N)$ gauge theories with matter in two-index representations,
i.e.~$\Ysymm$ (symmetric) or $\Yasymm$ (antisymmetric). In this section we
briefly discuss these two cases.

\subsection{{\large $\SU(N)$} with matter in the $\protect\Ysymm$ representation }
\label{sSUS}
Going through the same steps as for the previously discussed cases one may derive
the contour integral
\ben
Z_k &= &{1 \over k!} {\epsilon^k\over (\epsilon_1 \epsilon_2)^k}
\frac{1}{(2\pi i)^k} \oint \prod_{I=1}^k \D\phi_I
{ (2\phi_I{+}\ep_1)(2\phi_I{+}\ep_2)\prod_{i=1}^{N} (\phi_I +a_i)
\over P(\phi_I) P(\phi_I + \epsilon)} \non \\
&\times & \!\!\! \prod_{I<J}
\frac{(\phi_I-\phi_J)^2 ((\phi_I-\phi_J)^2-\ep^2) (\phi_I +\phi_J+\ep_1)
(\phi_I+\phi_J+\ep_2)}
{((\phi_I-\phi_J)^2-\ep_1^2)((\phi_I-\phi_J)^2-\ep_2^2)
(\phi_I+\phi_J)(\phi_I+\phi_J+\ep)}\,.
\label{susintegral}
\een
This case was briefly mentioned in \cite{Nekrasov:2002b}\footnote{There is a typo
in the formulas given in this reference. The unpublished corrected version
agrees with our expression. We thank S. Naculich for correspondence on this point.}.
From the integral (\ref{susintegral}) it is easy to see (assuming that the
contour is the same as in the case without matter) that
the solutions which contribute are the same ones as in the pure $\SU(N)$ theory.
Using this result one may derive  
\bea
Z_{\mathbf{k}} &=& 4^k \prod_{(i,j)\in Y_{\ell}}
[a_{\ell} + \hbar(i - j - \half)]^{1/2}
[a_{\ell} + \hbar(i - j)] [a_{\ell} + \hbar(i - j + \half)]^{1/2}   \\
& \times & \prod_{Y_{\ell}, Y_{r}} \prod_{m,n=1}^{\infty}
\frac{[a_{\ell} - a_{r} + \hbar(k_{\ell,m} - k_{r,n} +n - m)][a_{\ell} + a_{r}
+ \hbar(n - m)]^{1/2}}{ [a_{\ell} - a_{r} + \hbar(n - m)]
[a_{\ell} + a_{r} + \hbar(k_{\ell,m} - \tilde{k}_{r,n} +n - m)]^{1/2} }
 \non
\eea
where $\hbar = \ep_2 = -\ep_1$. In this expression it is assumed that the
points $(\ell,m)=(r,n)$ are excluded.

We have checked that the above expressions agree (up to two instantons)
with the results obtained in~\cite{Gomez-Reino:2003}
(see appendix C for details).

It should also be possible to analyze this model along the lines 
of \cite{Nekrasov:2003} and in particular derive the cubic curve (\ref{suscurve}).
As a first step we write down the analogue of (\ref{fullZ}):
\bea 
\mathcal{Z}_{\bf k} &\!\!=&\!\! \exp\bigg\{ -{\ts \frac{1}{4}} \bint f_{{\bf k},a}''(x)f''_{{\bf k},a}(y)\ga_{\hbar}(x-y,\La)
+{\ts \frac{1}{8}} \int  f_{{\bf k},a}''(x)f''_{{\bf k},a}(y)\ga_{\hbar}(x+y,\La) \non \\ &&
\qquad \; + \, {\ts \frac{1}{4}} \int f_{{\bf k},a}''(x) [ 2 \ga_{\hbar}(x,\La) + \ga_{\hbar}(x+\hbar/2,\La) + \ga_{\hbar}(x-\hbar/2,\La) ] \bigg\}
\eea
which leads to the following analogue of (\ref{pertZ})
\bea 
Z_{\mathrm{pert}} &\!\!=&\!\! \exp\bigg\{ -\sum_{\ell\neq r}\ga_{\hbar}(a_\ell-a_r,\La)
+\half \sum_{\ell, r} \ga_{\hbar}(a_\ell+a_r,\La) \non \\ && \qquad\, +\,
\half \sum_{\ell} 
[ 2 \ga_{\hbar}(a_\ell,\La) + \ga_{\hbar}(a_{\ell}+\hbar/2,\La) + \ga_{\hbar}(a_\ell-\hbar/2,\La) ] 
\bigg\} \,.
\eea
From which one can extract, using (\ref{ga0}),
\be
 \cF^{\mathrm{pert}}_{0} =- \sum_{\al\in\Delta_{+}} (a \cdot \al)^2 
\ln\bigg(\frac{a\cdot\al}{\La}\bigg) + \half \sum_{\mu \in R_{W}} (a \cdot \mu)^2 
\ln\bigg(\frac{a\cdot\mu}{\La}\bigg) + \mathrm{quadratic}\,,
\ee
where $\Delta_{+}$ is the set of positive roots of $\SU(N)$ and $R_{W}$ is the set of 
weights of the symmetric representation. This result agrees with the known result.

\subsection{{\large $\SU(N)$} with matter in the $\protect\Yasymm$ representation }
\label{sSUA}
For this case one finds the contour integral
\ben
Z_k &= &{1 \over k!} {\epsilon^k\over (\epsilon_1 \epsilon_2)^k}
\frac{1}{(2\pi i)^k} \oint \prod_{I=1}^k \D\phi_I
{ \prod_{i=1}^{N} (\phi_I +a_i)
\over P(\phi_I) P(\phi_I + \epsilon)(2\phi_I)(2\phi_I+\ep)} \non  \\
&\times & \!\!\! \prod_{I<J}
\frac{(\phi_I-\phi_J)^2 ((\phi_I-\phi_J)^2-\ep^2) (\phi_I +\phi_J+\ep_1)
(\phi_I+\phi_J+\ep_2)}
{((\phi_I-\phi_J)^2-\ep_1^2)((\phi_I-\phi_J)^2-\ep_2^2)
(\phi_I+\phi_J)(\phi_I+\phi_J+\ep)}\,.
\label{suaintegral}
\een
For this model one encounters the problem that in addition to the fixed points of 
the pure $\SU(N)$ theory extra solutions to the fixed point equations 
need to be taken into account to get
agreement with previous results. This is similar to the situation that
occurred for $\Sp(2N)$, which should come as no surprise given the similarity of
the Seiberg-Witten curves and the results
in~\cite{Kraus:2003,Naculich:2003b}. We have been
unable to find a closed expression which includes the extra contributions. Instead we
have checked that the above integral leads to results (up to two-instanton order) 
which are in agreement with the results~\cite{Gomez-Reino:2003} obtained 
using the Seiberg-Witten procedure.

\setcounter{equation}{0}
\section{Discussion}

Clearly there are several issues one would like to understand better. One problem
is to determine the correct stability condition for the $\Sp(2N)$
and $\SU(N)$+$\Yasymm$ cases and use it to explicitly evaluate the
contour integrals by summing up the residues.

Another problem concerns the relation with topological strings.
In the case of $\SO(2N)$ theories, one can in principle compute the
prepotential of the 5d theory on $\R^4 \times {\bf S}^1$, following
the same steps as in \cite{Nekrasov:2002}.
The resulting expression might be related to a topological string amplitude
on the Calabi-Yau obtained by fibering a $D_N$ singularity over a $\PP^1$ base.
This would be very interesting since no results are known for such manifolds, but
unfortunately there seems to be a conundrum: one would expect it to be possible
to write the topological string amplitude in terms of the K\"ahler parameters 
of the Calabi-Yau.
These are presumably in a one-to-one correspondence with the simple roots of
$D_{N}$, but the five-dimensional prepotential seems to involve combinations of the
K\"ahler parameters that do not correspond to sums of simple roots with positive
coefficients. It would be
interesting to clarify this and in that way obtain predictions for the
topological string amplitudes.

The methods we used in this paper can
also be applied to a similar problem, the determination of the so called
bulk (or principal) contribution to the Witten index in SYM quantum mechanics. 
It is known that this quantity can be written
as a so called Yang-Mills integral which can be reduced to 
a contour integral using methods similar
to the ones discussed in this paper (see e.g.~\cite{Moore:1998,Austing:2001} and
references therein for further details). Explicit expressions for the bulk part of
the Witten index are known only for the case of $\SU(N)$ (see e.g.~\cite{Moore:1998}
and references therein). However for the other classical gauge groups much less is
known and the results in the literature have been obtained
order-by-order using computer assisted
calculations~\cite{Krauth:2000}. 
The bulk Witten index can be studied for SYM quantum mechanics  obtained by 
dimensional reduction of $d=4$, $d=6$ and $d=10$ supersymmetric Yang-Mills theories. 
Our analysis corresponds most closely to the $d=6$ case.
As an example, for the $d=6$ $\Sp(2N)$ case the fixed points can be obtained by solving
\be
\ba{lll}
[\phi,B_1]=\ep_1 B_1  \,,&  [\phi,B_2] = \ep_2 B_2  \,, &  [B_1,B_2] = 0\,, \\
\phi^T g = - g\, \phi \,, &  B_1^T g = - g B_1  \,, &  B_2^T g = - g B_2 \,,\\
g^T = - g  \,, &  g \, g^* = -\1_{2N} \,.&
\ea
\ee
These equations are similar to the ones we solved in section \ref{sSp} and again
the solution can be represented pictorially in terms of stacks of row of boxes.
In addition to the rules we listed in section \ref{sSp} we now have the additional restrictions (related to the fact that the equations are not quite the same):
\begin{itemize}
\item Only rows containing an even number of boxes can be placed on top of 
the $x$-axis. For example, diagrams containing only one row can only have an even number of boxes.
\item Only an odd number of boxes are allowed to touch each other along the
$x$-axis.
\end{itemize}

These rules (together with the ones in section \ref{sSp}) give the solutions to the fixed-point equations which do not have moduli. Thus we have determined the possible locations of poles in the integrand of the contour integral which calculates the bulk part of the Witten index for the $\Sp(2N)$ quantum mechanics arising from $d=6$. However, just as in section \ref{sSp} not all of these solutions are relevant to the evaluation of the integral; what is lacking is an understanding of the stability condition. It is our hope that our results will be of some help in resolving the longstanding problem of calculating the bulk Witten index for SO/Sp supersymmetric quantum mechanics.

One can also apply our methods to cases other than $d=6$. For instance, for the
$\Sp(2N)$ ($\SO(N)$) theory arising from $d=4$ the solutions to the fixed point 
equations are given by direct sums of distinct even-dimensional (odd-dimensional) 
representations of $\SU(2)$, but once again the 
correct stability condition is not known.

\section*{Acknowledgements}
We would like to thank Ugo Bruzzo, Jos\'e Edelstein, 
Albrecht Klemm, Steve Naculich, and Nikita Nekrasov for 
useful conversations.

\appendix
\section*{Appendices}

\sectiono{{\large $\SO(N)$} instanton corrections up to {\large $k=3$}}

Here we discuss the explicit expressions which result from the formul\ae{} in
section \ref{sSO} and check that they agree with known results for $k=1,2,3$.
The contribution of each fixed point is obtained by evaluating the residue
of the integrand
in (\ref{sonintegral}) at the poles (\ref{fixedson}). Since the poles are
as in the $\U(N)$
case, the calculation is essentially the same as in that case.
 The result for $\epsilon_1=-\epsilon_2
=\hbar$ can be written in terms
of the function
\be
U_l (x)= {1 \over \prod_{m\not=l} ((a_{l}-a_m)^2 + x)^2}
{16(a_l+x)^4\over \prod_m (a_l+a_m +x)^2}
\ee
for $\SO(2N)$, and
\be
U_l (x)= {1 \over \prod_{m\not=l} ((a_{l}-a_m)^2 + x)^2}
{16(a_l+x)^2\over \prod_m (a_l+a_m +x)^2}
\ee
for $\SO(2N+1)$. Below we will use the notation $U_l=U_l(0)$,
$U^{(n)}_l=(\partial U_l (x)/\partial x)_{x=0}$.
After a simple (but tedious) calculation one obtains the first few
instanton corrections in the
$\SO(N)$ case: 
\ben
\hbar^2 Z_1 & \!\!\!\!\!\! =& \!\!\!\!\!\!\sum_{l} U_l \nonumber\\
\hbar^4 Z_2 &\!\!\!\!\!\! =&{ \!\!\!\!\!\! \ts \frac{1}{4}} \! \sum_{l=1}^m [U_l U_l(\hbar)
g^{(1)}_l(\hbar) {+} U_l U_l(-\hbar) g^{(1)}_l(-\hbar)]
{+}  \half \sum_{l\not = m}  U_l U_m g(a_{l}{-}a_m)g(a_l{+} a_m) \nonumber\\
\hbar^6 Z_3 &\!\!\!\!\!\!=&  \!\!\!\!\!\! {\ts \frac{1}{36}} \sum_l U_l \bigg[ U_l(\hbar) U_l(2\hbar)
g^{(2)}_l(\hbar)
+ U_l(-\hbar) U_l(-2\hbar) g^{(2)}_l(-\hbar) \non \\
&& \qquad \quad + \, 4 U_l(\hbar) U_l(-\hbar) g^{(3)}(\hbar) \bigg]   \\
&+& \!\!\!\! {\ts \frac{1}{4}} \sum_{l \not= m} U_l U_m \bigg[ U_l(\hbar)m(\hbar, a_{l}-a_{m} )
m(\hbar, a_l+a_m) g^{(1)}(\hbar) \non  \\
& & \qquad \qquad  + \,U_l(-\hbar)m( -\hbar, a_{l}-a_{m}) m(-\hbar, a_l+a_m)
g^{(1)}(-\hbar) \bigg] \non \\
&\!\!\!\!+& \!\!\!\! {\ts \frac{1}{6}} \!\!\!  \sum_{l\not= m\not= n} \!\!\!
U_l U_m U_n g(a_{l}{-}a_{m})g(a_{l}{-}a_{n})g(a_{m}{-}a_{n}) g(a_l{+}a_m)
g(a_l{+}a_n)g(a_m{+}a_n) \non
\een
where we have used the notation
\ben
m(x,y)& = &{1\over \Bigl[ 1-{2x^2 \over y (x+y)}\Bigr]^2}\,, \qquad
g(y) ={1\over \Bigl( 1 -{\hbar^2\over y^2}\Bigr)^2}\,, \qquad
g^{(1)}_l(x) = {1\over\Bigl[ 1 -\bigl({x\over 2 a_l +x}\bigr)^2\Bigr]^2}\,,
\nonumber\\
g^{(2)}_l(x)&=&{1\over\Bigl[ 1 -\bigl({x\over 2a_l +x}\bigr)^2\Bigr]^2\Bigl[ 1 -\bigl({x\over 2a_l +2 x}\bigr)^2\Bigr]^2
\Bigl[ 1 -\bigl({x\over 2a_l +3x}\bigr)^2\Bigr]^2}\,,\nonumber\\
 g^{(3)}_l(x)&=&{1\over\Bigl[ 1 -\bigl({x\over 2a_l +x}\bigr)^2\Bigr]^2\Bigl[ 1 -\bigl({x\over 2a_l - x}\bigr)^2\Bigr]^2
\Bigl[ 1 -\bigl({x\over 2a_l})^2\Bigr]^2}\,.
\een
From these equations one can compute the instanton corrections to the prepotential,
$F_k\equiv F_{0,k}$ for $k=1,2,3$. The results are as follows:
\ben
F_1&=& \sum_{l} U_l, \nonumber\\
F_2&=& {\ts \frac{1}{4}} \sum_{l}U_l  \bigl( U_l^{''} {+} {U_l\over a_l^2}\bigr) +
 \sum_{l\not = m}  U_l U_m \biggl( {1\over (a_{l}-a_m)^2} + {1 \over (a_l+a_m)^2}\biggr),\nonumber\\
F_3&=& {\ts \frac{1}{36}} \sum_l U_l[U_l U_l^{(4)} {+} 2 U_l U_l^{'''} {+} 3(U_l^{''})^2]
{+} {\ts \frac{1}{16}} \sum_l {1\over a_l^4}(5 U_l^3 {-} 4 a_l U_l^2 U_l' {+} 4 U_l^2 U_l^{''})
\nonumber\\
&+& \sum_{l\neq m} U_lU_m\Biggl\{ 5U_l \biggl( {1 \over (a_{l}{-}a_m)^4} + {1 \over (a_l {+} a_m)^4} \biggr)
 {-} 2U_l^{'} \biggl( {1 \over (a_{l}{-}a_m)^3} + {1 \over (a_l {+} a_m)^3} \biggr)
\non \\
&& \qquad +\, U_l^{''} \biggl( {1 \over (a_{l}-a_m)^2} + {1 \over (a_l + a_m)^2}  \biggr)
\Biggr\}
\non \\
&+& \sum_{l\neq m} U_l^2 U_m\Biggl\{ {1\over a_l^2}\biggl( {1 \over (a_{l}-a_m)^2} + {1 \over (a_l + a_m)^2} \biggr) +
{4 \over (a_{l}-a_m)^2 (a_l + a_m)^2}\Biggr\}
\nonumber\\
&+&{2} \sum_{l\not= m \not = n} U_l U_m U_n \Biggl\{ {1\over (a_{l}-a_m)^2 (a_{m}-a_n)^2} + {1\over (a_l+ a_m)^2 (a_m + a_n)^2}
\nonumber\\ & &
\qquad \qquad + \,2 {1\over (a_{l}-a_m)^2 (a_{n} + a_m)^2} \Biggr\}.
\een
We can compare these results to the ones obtained from the Seiberg-Witten approach
by utilizing the relation between the prepotential of the $\SO(2N)$ theory and
that of the $\SU(2N)$ theory with 4 massless flavors~\cite{D'Hoker:1997b}.
Using this relation together with the $\SU$
results in~\cite{Chan:1999} (or \cite{Nekrasov:2002}) one finds agreement with
our results above after a rescaling $F_k \rightarrow 2^{-4k+1}F_k$.

\setcounter{equation}{0}
\section{{\large $\Sp(2N)$}: instanton corrections up to {\large $k=3$} }
Here we will perform a check of our results by comparing the first three
instanton corrections to the prepotential obtained using the Seiberg-Witten
approach to those obtained using many-instanton counting.

\medskip
{\it Seiberg-Witten approach}
\medskip

The Seiberg-Witten curve for the $\Sp(2N)$ gauge theory without matter is
(see e.g.~\cite{D'Hoker:1997b,Landsteiner:1998} and references therein)
\be \label{Spcurve}
y^2 + 2 \, y \, [ x^2\prod_{i=1}^{N}(x^2-e_i^2) + L] + L^2 =0 \,,
\ee
where $L = \La^{2N+2}$.
The quantum order parameters are \cite{D'Hoker:1997a}
\be
a_i = e_i + \sum_{\mbox{\tiny $\ba{c} m,n\ge0 \\ (m,n)\neq(0,0) \ea $}}
\frac{(-1)^n L^{n+2m}}{2^{2m} (m!)^2 n! }\bigg[ \bigg(\frac{\pa}{\pa x}\bigg)^{2m+n-1}R_k(x)^n S_k(x)^m \bigg]_{x=e_k}
\ee
where $S_k(x) \equiv (R_k(x))^2$ and
\be
\frac{1}{ x^2\prod_{i=1}^{N}(x^2-e_i^2)} = \frac{R_k(x)}{x-e_k} = \frac{R_0(x)}{x^2} \,.
\ee
For later reference we have also introduced $R_0(x)$ with $S_0(x) \equiv (R_0(x))^2$.
The instanton expansion of the prepotential can be obtained using the recursive
methods developed in~\cite{Chan:1999}, which leads to
(after using various
identities)\footnote{The first two expressions were first obtained in~\cite{Ennes:1999}.}
\bea  \label{Spinst}
-F_1 &=& R_0(0) \,,\non \\
-F_2 &=& -\half \sum_k S_k(a_k) - {\ts \frac{1}{8}} S_0''(0)\,, \non \\
-F_3 &=& R_0(0) \bigg( {\ts \frac{3}{2}}  \sum_k \frac{S_k(a_k)}{a_k^2}
+ {\ts \frac{1}{96}}  S_0^{(4)}(0)  \bigg) \,.
\eea

\medskip
{\it Many-instanton counting}
\medskip

From the contour integrals in section \ref{sSp} we find
\bea
\hbar^2 Z_1 &=& \frac{1}{2P(0)} \,, \non \\
\hbar^4 Z_2 &=& \frac{1}{8} \bigg[ \frac{1}{P(\hbar/2)^2}
+ \frac{\hbar^2}{8} \sum_k \frac{1}{(a_k^2-(\hbar/2)^2)^2a_k^2
\prod_{j\neq k} (a_k^2-a_j^2)^2} \bigg] \,,\non \\
\hbar^6 Z_3 &=& \frac{1}{16}  \frac{1}{P(0)} \Bigg[
\frac{2}{9} \frac{1}{P(\hbar)^2} + \frac{1}{9} \frac{1}{P(\hbar/2)^2}
 \non \\
&& \qquad  + \, \frac{\hbar^2}{8} \sum_k \frac{a_k^2}{(a_k^2-\hbar^2)^2
(a_k^2-(\hbar/2)^2)^2 \prod_{j\neq k} (a_k^2-a_j^2)^2} \Bigg]\,.
\eea
Using the relations 
\be \label{FZrel}
F_1 = \hbar^2 Z_1 \,, \qquad F_2 = \hbar^2(Z_2 - \frac{Z_1^2}{2}) \,,
\qquad F_3 = \hbar^2(Z_3 - Z_1 Z_2 + \frac{Z_1^3}{3}) \,,
\ee
and extracting the $\hbar$-independent pieces we find agreement with the
above results (\ref{Spinst})  obtained from the Seiberg-Witten procedure
provided that $F_k^{\rm here} = - \frac{F_k^{\rm there}}{k\, 2^k}$. Part of this difference
can be removed by a rescaling of $\La$; the other part is a result of
different conventions.

As a further check we have also verified that the above results are consistent
with the ones for $\SO(2N+1)$ using the isomorphism $\SO(5)\cong \Sp(4)$
and allowing for a rescaling of the $\La$'s in the two theories.
(The translation between the order parameters, $d_i$, of $\SO(5)$ and the
ones of $\Sp(4)$, $a_i$, are:
$d_1 = a_1-a_2$ and $d_2 = a_1+a_2$.)

\setcounter{equation}{0}
\section{{\large $\SU(N)\,+\,$}$\protect\Yasymm$/$\protect\Ysymm$: instanton corrections up to {\large $k=2$}  }

Here we will perform a check of our results for the
$\SU(N)\,+\,\Yasymm$/$\Ysymm$ theories by comparing the first two
instanton corrections to the prepotential obtained using the Seiberg-Witten
approach to those obtained from many-instanton counting.

\subsection{{\large $\SU(N)\,+\,\protect\Ysymm$}}

\medskip
{\it Seiberg-Witten approach}
\medskip

The cubic Seiberg-Witten curve for the $\SU(N)$ gauge
theory with one matter hypermultiplet in the symmetric ($\Ysymm$) representation
is given by \cite{Landsteiner:1998}
\be \label{suscurve}
y^3 + P(x) y^2 + x^2 P(-x) L + x^6 L^3 = 0\,,
\ee
where $P(x) = \prod_{i=1}^N (x-e_i)$ and $L=\La^{N-2}$. The instanton expansion
of the prepotential has been obtained to the first few orders 
in~\cite{Gomez-Reino:2003}\footnote{The
one-instanton expression was first obtained in \cite{Naculich:1998}.}. 
After using various identities
one finds the expressions
\bea \label{susinst}
-F_1 &=& - \sum_k S_k(a_k)  \,, \\
-F_2 &=&   {\ts \frac{1}{4} } \sum_k S_k(a_k)S_k''(a_k)
+ \sum_{k\neq l} \frac{S_k(a_k)S_l(a_l)}{(a_l-a_k)^2}
- \half  \sum_{k, l} \frac{S_k(a_k)S_l(a_l)}{(a_l+a_k)^2} \,, \non
\eea
where we have used the definition
\be
S_k(x) = \frac{x^2 \prod_{i}(-x-a_i)}{\prod_{j\neq k} (x-a_j)^2}\,.
\ee

\medskip
{\it Many-instanton counting}
\medskip

From the expressions in section \ref{sSUS} we find
\bea
\hbar^2 Z_1 &=& -\,4 \sum_k (a_k^2 - (\hbar/2)^2)\, 
\frac{\prod_{j}(a_k+a_j)}{\prod_{j \neq k} (a_k-a_j)^2}
 \\
\hbar^4 Z_2 &=& 16 \bigg[  {\ts \frac{1}{4}} \sum_k (a_k - 3 \hbar/2)\, a_k\,
(a_k - \hbar)(a_k + \hbar/2)
\frac{ \prod_{j}(a_k-\hbar+a_j) (a_k+a_j)}{\prod_{j \neq k}(a_k-\hbar-a_j)^2
(a_k{-}a_j)^2 }  \non \\
&& +\, {\ts \frac{1}{4}} \sum_k (a_k + 3 \hbar/2)\, a_k \,
(a_k + \hbar)(a_k - \hbar/2)
\frac{ \prod_{j}(a_k+\hbar+a_j) (a_k+a_j)}{\prod_{j \neq k}(a_k+\hbar-a_j)^2
(a_k-a_j)^2 }  \non \\
&&\!\!\!\!\!\!\!\!\!\! +\, \half \sum_{k\neq l} \frac{(a_k^2 {-} (\hbar/2)^2)
(a_l^2 {-} (\hbar/2)^2) (a_k-a_l)^4
((a_k {+} a_l)^2{-}\hbar^2) \prod_{j}(a_k{+}a_j)(a_l{+}a_j) }
{((a_l-a_k)^2-\hbar^2)^2 (a_k+a_k)^2 \prod_{j \neq k} (a_k-a_j)^2
\prod_{j\neq l}(a_l-a_j)^2}   \bigg] . \non
\eea

Using (\ref{FZrel}) and extracting the leading pieces, we find agreement with
the results in (\ref{susinst})
provided that we identify $F_k^{\rm here} = (-4)^k (-1)^{Nk} F_{k}^{\rm there} $.

\subsection{{\large $\SU(N)\,+\,\protect\Yasymm$}}

\medskip
{\it Seiberg-Witten approach}
\medskip

The non-hyperelliptic Seiberg-Witten curve for the $\SU(N)$ gauge
theory with one matter hypermultiplet in the antisymmetric ($\Yasymm$) representation
is given by \cite{Landsteiner:1998}
\be
y^3 + \bigg[P(x)+\frac{3L}{x^2}\bigg] y^2 + \frac{L}{x^2} [P(-x)+\frac{3L}{x^2}\bigg]y  +\frac{L^3}{x^6}  = 0 \,,
\ee
where $P(x) = \prod_{i=1}^N (x-e_i)$ and $L=\La^{N+2}$. The instanton expansion
of the prepotential has been obtained to the first few orders 
in~\cite{Gomez-Reino:2003}\footnote{The
one-instanton expression was first obtained in \cite{Ennes:1998}.}. 

After using various identities one obtains
\bea \label{suainst}
F_1 &=& \sum_k S_k(a_k)  - 2 S_0(0) \,,\\
F_2 &=&  {\ts \frac{1}{4} } \sum_k S_k(a_k)S_k''(a_k)
+ \sum_{k\neq l} \frac{S_k(a_k)S_l(a_l)}{(a_l-a_k)^2}
- \half  \sum_{k, l} \frac{S_k(a_k)S_l(a_l)}{(a_l+a_k)^2} \non \\
&& -\, 2 S_0(0) \sum_k \frac{S_k(a_k)}{a_k^2} + \half S_0(0)S_0''(0)
- \half S_0'(0)S_0'(0) \,, \non
\eea
where we have used the definitions
\be
\frac{\prod_{k}(-x-a_k)}{x^2 \prod_k (x-a_k)^2} \equiv \frac{S_k(x)}{(x-a_k)^2}
\equiv \frac{S_0(x)}{x^2}\,.
\ee

\medskip
{\it Many-instanton counting}
\medskip

Evaluation of the contour integrals gives
\bea \label{suaZ}
\hbar^2 Z_1 &=& - {\ts \frac{1}{4}} \sum_k
\frac{\prod_{j}(a_k+a_j)}{a_k^2\prod_{j \neq k} (a_k-a_j)^2}
- \half \frac{\prod_i a_i}{\prod_i (-a_i^2)} \,,
 \\
\hbar^4 Z_2 &=& {\ts \frac{1}{16} } \bigg[  {\ts \frac{1}{4} } \sum_k \frac{1}{(a_k - \hbar)\, a_k\, (a_k - \hbar/2)^2}
\frac{ \prod_{j}(a_k-\hbar+a_j) (a_k+a_j)}{\prod_{j \neq k}(a_k-\hbar-a_j)^2
(a_k{-}a_j)^2 }  \non \\
&& \;\, +\, {\ts \frac{1}{4}} \sum_k \frac{1}{(a_k + \hbar)\, a_k\, (a_k + \hbar/2)^2}
\frac{ \prod_{j}(a_k + \hbar+a_j) (a_k+a_j)}{\prod_{j \neq k}(a_k+\hbar-a_j)^2
(a_k{-}a_j)^2 }  \non \\
&& +\, \half \sum_{k\neq l} \frac{ (a_k-a_l)^4
((a_k +a_l)^2-\hbar^2) \prod_{j}(a_k+a_j)(a_l+a_j) }
{((a_l{-}a_k)^2-\hbar^2)^2 (a_k+a_k)^2 \,a_l^2 \,a_k^2 \prod_{j \neq k} (a_k-a_j)^2
\prod_{j\neq l}(a_l-a_j)^2}   \non \\
&&- 2\sum_k \frac{a_k^2}{ a_k^2-\hbar^2}
\frac{ \prod_{j}(a_j) (a_k+a_j)}{\prod_{j }(-a_j)^2
\prod_{i\neq k}(a_k{-}a_i)^2 } + 2\frac{1}{\prod_j (\hbar/2-a_j)(-\hbar/2-a_j))}
\bigg]  , \non
\eea
Using (\ref{FZrel}) and extracting the leading pieces we find agreement with
(\ref{suainst}) provided that we identify 
$F_k^{\rm here} = (-4)^{-k} (-1)^{Nk} F_{k}^{\rm there} $.

To arrive at the result (\ref{suaZ}) one needs to choose a contour 
which picks out a very particular set of poles. We do not understand 
why this particular contour should be chosen (knowledge of the correct 
stability condition should shed light on this). 

As a consistency check we have checked that the above results are consistent
with the fact that for $\SU(3)$ the antisymmetric representation is isomorphic
with the fundamental representation\footnote{One can also make the comparison 
for $\U(3)$ but in that case one also needs to shift $a_i \rar a_i -(a_1+a_2+a_3)/2$.}.

\begingroup\raggedright\endgroup


\begin{thebibliography}{10}

\bibitem{Seiberg:1994a}
N.~Seiberg and E.~Witten, ``Electric-magnetic duality, monopole condensation,
  and confinement in $\mathcal{N}=2$ supersymmetric Yang-Mills theory,'' {\em
  Nucl. Phys.} {\bf B426} (1994) 19, {{\tt hep-th/9407087}};
N.~Seiberg and E.~Witten, ``Monopoles, duality and chiral symmetry breaking in
  $\mathcal{N}=2$ supersymmetric QCD,'' {\em Nucl. Phys.} {\bf B431} (1994)
  484, {{\tt hep-th/9408099}}.

\bibitem{Nekrasov:2002}
N.~A. Nekrasov, ``Seiberg-Witten prepotential from instanton counting,''
{{\tt hep-th/0206161}}.

\bibitem{Dorey:2002}
N.~Dorey, T.~J. Hollowood, V.~V. Khoze, and M.~P. Mattis, ``The calculus of
  many instantons,'' {\em Phys. Rept.} {\bf 371} (2002) 231,
{{\tt hep-th/0206063}}.

\bibitem{Losev:1997}
A.~Losev, N.~Nekrasov, and S.~L. Shatashvili, ``Testing Seiberg-Witten
  solution,'' {{\tt hep-th/9801061}}.

\bibitem{Klemm:2002}
A.~Klemm, M.~Mari\~no, and S.~Theisen, ``Gravitational corrections in
  supersymmetric gauge theory and matrix models,'' {\em JHEP} {\bf 03} (2003)
  051,
{{\tt hep-th/0211216}}.

\bibitem{Iqbal:2003a}
A.~Iqbal and A.-K. Kashani-Poor, ``Instanton counting and Chern-Simons
  theory,'' {\em Adv. Theor. Math. Phys.} {\bf 7} (2004) 457,
{{\tt hep-th/0212279}};
A.~Iqbal and A.-K. Kashani-Poor, ``$\SU(N)$ geometries and topological string
  amplitudes,''
{{\tt hep-th/0306032}};
T.~Eguchi and H.~Kanno, ``Topological strings and Nekrasov's formulas,'' {\em
  JHEP} {\bf 12} (2003) 006,
{{\tt hep-th/0310235}}.

\bibitem{Flume:2002}
R.~Flume and R.~Poghossian, ``An algorithm for the microscopic evaluation of
  the coefficients of the Seiberg-Witten prepotential,'' {\em Int. J. Mod.
  Phys.} {\bf A18} (2003) 2541, {{\tt hep-th/0208176}}.

\bibitem{Bruzzo:2002}
U.~Bruzzo, F.~Fucito, J.~F. Morales, and A.~Tanzini, ``Multi-instanton calculus
  and equivariant cohomology,'' {\em JHEP} {\bf 05} (2003) 054, {{\tt hep-th/0211108}}.

\bibitem{Losev:2003}
A.~S. Losev, A.~Marshakov, and N.~A. Nekrasov, ``Small instantons, little
  strings and free fermions,''
{{\tt hep-th/0302191}}.

\bibitem{Nakajima:2003a}
H.~Nakajima and K.~Yoshioka, ``Instanton counting on blowup. I,''
{{\tt math.ag/0306198}}.

\bibitem{Nekrasov:2003}
N.~Nekrasov and A.~Okounkov, ``Seiberg-Witten theory and random partitions,''
{{\tt hep-th/0306238}}.

\bibitem{Hollowood:2003}
T.~J. Hollowood, A.~Iqbal, and C.~Vafa, ``Matrix models, geometric engineering
  and elliptic genera,''
{{\tt hep-th/0310272}}.

\bibitem{Nakajima:2003b}
H.~Nakajima and K.~Yoshioka, ``Lectures on instanton counting,''
{{\tt math.ag/0311058}}.

\bibitem{Flume:2004}
Y.~Tachikawa,
``Five-dimensional Chern-Simons terms and Nekrasov's instanton counting,''
{\em JHEP} {\bf 02} (2004) 050,
{\tt hep-th/0401184}; \\
R.~Flume, F.~Fucito, J.~F. Morales, and R.~Poghossian, ``Matone's relation in
  the presence of gravitational couplings,''
{{\tt hep-th/0403057}}.

\bibitem{D'Hoker:1997b}
E.~D'Hoker, I.~M. Krichever, and D.~H. Phong, ``The effective prepotential of
  $\mathcal{N} = 2$ supersymmetric $\SO(N_c)$ and $\Sp(N_c)$ gauge
  theories,'' {\em Nucl. Phys.} {\bf B489} (1997) 211,
{{\tt hep-th/9609145}}.

\bibitem{Naculich:1998}
S.~G. Naculich, H.~Rhedin, and H.~J. Schnitzer, ``One-instanton test of a
  Seiberg-Witten curve from M-theory: the antisymmetric representation of
  $\SU(N)$,'' {\em Nucl. Phys.} {\bf B533} (1998) 275,
{{\tt hep-th/9804105}}.

\bibitem{Ennes:1999}
I.~P. Ennes, C.~Lozano, S.~G. Naculich, and H.~J. Schnitzer, ``Elliptic models
  and M-theory,'' {\em Nucl. Phys.} {\bf B576} (2000) 313,
{{\tt hep-th/9912133}}.

\bibitem{Dijkgraaf:2002}
R.~Dijkgraaf and C.~Vafa, ``A perturbative window into non-perturbative
  physics,''
{{\tt hep-th/0208048}}.

\bibitem{Kraus:2003}
P.~Kraus and M.~Shigemori, ``On the matter of the Dijkgraaf-Vafa conjecture,''
  {\em JHEP} {\bf 04} (2003) 052,
{{\tt hep-th/0303104}};
F.~Cachazo, ``Notes on supersymmetric $\Sp(N)$ theories with an
  antisymmetric tensor,''
{{\tt hep-th/0307063}};
K.~Intriligator, P.~Kraus, A.~V. Ryzhov, M.~Shigemori, and C.~Vafa, ``On low
  rank classical groups in string theory, gauge theory and matrix models,''
{{\tt hep-th/0311181}}.

\bibitem{Naculich:2003b}
S.~G. Naculich, H.~J. Schnitzer, and N.~Wyllard, ``Matrix-model description of
  $\mathcal{N} = 2$ gauge theories with non-hyperelliptic Seiberg-Witten
  curves,'' {\em Nucl. Phys.} {\bf B674} (2003) 37,
{{\tt hep-th/0305263}}.


\bibitem{Atiyah:1978}
M.~F. Atiyah, N.~J. Hitchin, V.~G. Drinfeld, and Y.~I. Manin, ``Construction of
  instantons,'' {\em Phys. Lett.} {\bf A65} (1978)
185.

\bibitem{Witten:1996}
E.~Witten, ``Small instantons in string theory,'' {\em Nucl. Phys.} {\bf B460}
  (1996) 541,
{{\tt hep-th/9511030}};
M.~R. Douglas, ``Gauge fields and D-branes,'' {\em J. Geom. Phys.} {\bf 28}
  (1998) 255,
{{\tt hep-th/9604198}};
M.~R. Douglas, ``Branes within branes,''
{{\tt hep-th/9512077}}.

\bibitem{Nakajima:1999}
H.~Nakajima, {\em Lectures on Hilbert schemes of points on surfaces}, vol.~18
  of {\em University lecture series}.
\newblock American Mathematical Society, 1999.

\bibitem{Corrigan:1978}
E.~Corrigan, D.~B. Fairlie, S.~Templeton, and P.~Goddard, ``A Green's function
  for the general selfdual gauge field,'' {\em Nucl. Phys.} {\bf B140} (1978)
31;
N.~H. Christ, E.~J. Weinberg, and N.~K. Stanton, ``General self-dual Yang-Mills
  solutions,'' {\em Phys. Rev.} {\bf D18} (1978)
2013.

\bibitem{Gimon:1996}
E.~G. Gimon and J.~Polchinski, ``Consistency conditions for orientifolds and
  D-manifolds,'' {\em Phys. Rev.} {\bf D54} (1996) 1667,
{{\tt hep-th/9601038}}.

\bibitem{Hollowood:2002a}
T.~J. Hollowood, ``Calculating the prepotential by localization on the moduli
  space of instantons,'' {\em JHEP} {\bf 03} (2002) 038,
{{\tt hep-th/0201075}};
T.~J. Hollowood, ``Testing Seiberg-Witten theory to all orders in the instanton
  expansion,'' {\em Nucl. Phys.} {\bf B639} (2002) 66,
{{\tt hep-th/0202197}}.

\bibitem{Moore:1998}
G.~W. Moore, N.~Nekrasov, and S.~Shatashvili, ``D-particle bound states and
  generalized instantons,'' {\em Commun. Math. Phys.} {\bf 209} (2000) 77,
{{\tt hep-th/9803265}}.

\bibitem{Braverman:2004}
A.~Braverman, ``Instanton counting via affine Lie algebras I: equivariant
  j-functions of (affine) flag manifolds and Whittaker vectors,''
{{\tt math.ag/0401409}}.

\bibitem{Naculich:2001}
S.~G. Naculich, H.~J. Schnitzer, and N.~Wyllard, ``Vacuum states of
  $\mathcal{N} = 1*$ mass deformations of $\mathcal{N} = 4$ and $\mathcal{N} =
  2$ conformal gauge theories and their brane interpretations,'' {\em Nucl.
  Phys.} {\bf B609} (2001) 283,
{{\tt hep-th/0103047}}.

\bibitem{Dijkgraaf:2002b}
R.~Dijkgraaf, A.~Sinkovics, and M.~Temurhan, ``Matrix models and gravitational
  corrections,''
{{\tt hep-th/0211241}}.

\bibitem{emm}
J.~D.~Edelstein, M.~Mari\~no and J.~Mas,
``Whitham hierarchies, instanton corrections and soft supersymmetry breaking
in $\cN = 2$ $\SU(N)$ super Yang-Mills theory,''
{\em Nucl. Phys.} {\bf B541} (1999) 671, 
{\tt hep-th/9805172}; J.~D.~Edelstein, M.~G\'omez-Reino and J.~Mas,
``Instanton corrections in $\cN = 2$ supersymmetric theories with classical gauge
groups and fundamental matter hypermultiplets,''
{\em Nucl. Phys.} {\bf B561} (1999) 273, 
{\tt hep-th/9904087}.



\bibitem{Nekrasov:2002b}
N.~Nekrasov, ``Solution of $\mathcal{N}=2$ gauge theories via instanton
  counting.'' talk at Strings 2002, {\tt
  http://www.damtp.cam.ac.uk/strings02/avt/nekrasov/}

\bibitem{Gomez-Reino:2003}
M.~G\'omez-Reino, ``Prepotential and instanton corrections in $\mathcal{N} = 2$
  supersymmetric $\mathrm{SU}(N_1){\times}\mathrm{SU}(N_2)$ Yang-Mills
  theories,'' {\em JHEP} {\bf 03} (2003) 043,
{{\tt hep-th/0301105}}.

\bibitem{Austing:2001}
P.~Austing, ``Yang-Mills matrix theory,''
{{\tt hep-th/0108128}}.

\bibitem{Krauth:2000}
W.~Krauth and M.~Staudacher, ``Yang-Mills integrals for orthogonal, symplectic
  and exceptional groups,'' {\em Nucl. Phys.} {\bf B584} (2000) 641,
{{\tt hep-th/0004076}};
M.~Staudacher, ``Bulk Witten indices and the number of normalizable ground
  states in supersymmetric quantum mechanics of orthogonal, symplectic and
  exceptional groups,'' {\em Phys. Lett.} {\bf B488} (2000) 194,
{{\tt hep-th/0006234}};
V.~Pestun, ``$\cN = 4$ SYM matrix integrals for almost all simple gauge groups
  (except $E_7$ and $E_8$),'' {\em JHEP} {\bf 09} (2002) 012,
{{\tt hep-th/0206069}};
T.~Fischbacher, ``Bulk Witten indices from $d = 10$ Yang-Mills integrals,''
{{\tt hep-th/0312262}}.

\bibitem{Chan:1999}
G.~Chan and E.~D'Hoker, ``Instanton recursion relations for the effective
  prepotential in $\mathcal{N} = 2$ super Yang-Mills,'' {\em Nucl. Phys.} {\bf
  B564} (2000) 503,
{{\tt hep-th/9906193}}.

\bibitem{Landsteiner:1998}
K.~Landsteiner and E.~L\'opez, ``New curves from branes,'' {\em Nucl. Phys.} {\bf
  B516} (1998) 273,
{{\tt hep-th/9708118}}.

\bibitem{D'Hoker:1997a}
E.~D'Hoker, I.~M. Krichever, and D.~H. Phong, ``The effective prepotential of $\cN
  = 2$ supersymmetric $\SU(N_c)$ gauge theories,'' {\em Nucl. Phys.} {\bf B489}
  (1997) 179,
{{\tt hep-th/9609041}}.

\bibitem{Ennes:1998}
I.~P. Ennes, S.~G. Naculich, H.~Rhedin, and H.~J. Schnitzer, ``One instanton
  predictions of a Seiberg-Witten curve from M-theory: the symmetric
  representation of $\SU(N)$,'' {\em Int. J. Mod. Phys.} {\bf A14} (1999)
  301--321,
{{\tt hep-th/9804151}}.

\end{thebibliography}
\end{document}